\documentstyle{article}

\begin{document}

\title{ Fock Spaces with Reflection Condition and \\
Generalized Statistics}

\bigskip

\author{ Liu ZHAO\\
Institute of Modern Physics,  Northwest University, Xian 710069, China}

\newcommand{\bd}{\begin{displaymath}}
\newcommand{\ed}{\end{displaymath}}
\newcommand{\be}{\begin{equation}}
\newcommand{\ee}{\end{equation}}
\newcommand{\ba}{\begin{array}}
\newcommand{\ea}{\end{array}}
\newcommand{\bea}{\begin{eqnarray}}
\newcommand{\eea}{\end{eqnarray}}
\newcommand{\bee}{\begin{eqnarray*}}
\newcommand{\eee}{\end{eqnarray*}}
\renewcommand{\theequation}{\thesection.\arabic{equation}}
\def\p{\partial}
\def\alf{\alpha}
\def\bi{\beta}
\def\es{\epsilon}
\def\la{\lambda}
\def\dl{\delta}

\maketitle

\abstract{An oscillator algebra and the associated Fock space 
with reflecting boundary and generalized statistics are constructed 
and is generalized to the multicomponent case. 
The oscillator algebra depends manifestly
on the reflection factor and the statistical (exchange) factor,
and the corresponding Fock space can be obtained from that of 
the usual bosonic oscillator without reflection 
condition by certain projection operation.
}

\newpage

\section{Introduction}
\setcounter{equation}{0}

Oscillator algebras and their Fock representations play important basic 
roles in many branches of modern quantum field theory, condensed matter and 
solid state physics. Oscillator algebras are also the theoretical foundation 
of the second quantization method. In the recent years, as the theories of
quantum groups and quantum algebras develop, various generalized oscillator 
algebras have been proposed and acquired much attentions. Among these are 
$q$-oscillator algebra \cite{1,2}, $quon$-algebra \cite{3,4}, 
$(p,~q)$-deformed oscillator algebra \cite{5}-\cite{7} etc.

On the other hand, some hints show that the quasi-particles corresponding to 
the cluster excitations in solids, such as the anions that played some role 
in fractional quantum Hall effect, do not bey the usual Bose or Fermi 
statistics \cite{8}-\cite{11}. Following from this fact, the so-called 
Fock space and its associated oscillator algebra with generalized statistics 
are constructed \cite{12,13}. 

Recently, much progress has been made in the studies of massive field 
theoretical and lattice statistical models with integrable boundaries 
\cite{14,15}. One of the important progresses in these fields is the
establishment of the concept of ``boundary states" \cite{14,15}, 
which correspond to the lowest-energy states of quantum integrable field 
theories and/or solvable lattice statistical models with integrable 
boundaries. Based on this concept, systematical approaches for calculating
the boundary form factors and boundary correlation functions in some boundary 
solvable lattice statistical models \cite{15} and boundary integrable 
quantum field theories \cite{16} have been established. However, all these 
approaches use the technique of bosonization via (deformed) free fields 
and thus the generalizations to models with higher rank underlying 
(quantum) Lie algebras seems to be very complicated. 
Moreover, as the complete 
Fock space structure of boundary integrable field theories and solvable 
statistical models is not clear yet, the alternative approach (i.e. that
does not depend on the technique of bosonization) is difficult to be 
established.

In this article an oscillator algebra and the associated Fock space 
satisfying reflection boundary condition and obeying generalized statistics
will be constructed. The oscillator algebra depends crucially
on the boundary reflection factor while the corresponding Fock space 
can be obtained from that of the usual harmonic oscillator algebra 
with continuous spectrum via simple projection. The creation and anihilation 
operators in this new oscillator algebra is considered to correspond to 
quasi-particles in systems with generalized statistics, and the results 
can be generalized to the cases with more than one kinds of quasi-particle
excitations.

\section{Oscillator algebra with reflection boundary condition}
\setcounter{equation}{0}

Let us consider the continuum limit of an $s$-dimensional condensed 
matter system. In such a system the creation and anihilation operators 
of foundamental excitations (denoted respectively $b^\ast$ and $b$) will 
depend on a continuous $s$-vector $x$ which corresponds to the wave vector 
or spectrum of the excitations. In the case when there is no reflection 
boundary condition the creation and anihilation operators satisfy the 
following usual harmonic oscillator algebra,

\bea
& & [ b(x),~b(y) ] = [ b^\ast (x),~b^\ast (y) ] = 0, \nonumber\\
& & [ b(x),~b^\ast (y) ] = \delta (x - y), \label{1}
\eea

\noindent where $ \delta (x - y) = \prod_{i=1}^s~\delta ( x^i - y^i )$
and $x^i$ is the $i$-th component of $x$.

Now suppose that the system is bounded in some direction, say, in the first 
coordinate direction, and the action of the creation operator, say 
$a^\ast(x)$, on the 
lowest energy state $\Omega_B$ (the boundary vacuum state) satisfy the 
``reflection boundary condition''

\be
a^\ast (x) ~\Omega_B = K(x)~a^\ast (\sigma(x))~\Omega_B, \label{2}
\ee

\noindent where $\sigma(x)$ is the reflection of the ``wave vector'' $x$
which is defined as

\be
\sigma(x) \equiv \sigma(x^1,~x^2,~...,~x^s) = (-x^1,~x^2,~...,~x^s),
~~~\sigma^2 = Id,
\label{3}
\ee

\noindent $K(x)$ is called reflection factor which specify the status of 
the boundary. Requiring that Eq.(\ref{2}) be consistent, the following 
must hold for $K(x)$,

\be
K(x)~K(\sigma(x)) =1. \label{4}
\ee

\noindent 
In what follows we shall prove that the creation operator satisfying  
the conditions (\ref{2}-\ref{4}) and the anihilation operator which 
kills the vacuum state $\Omega_B$ satisfy an oscillator algebra which 
is different from eq.(\ref{1}), and the corresponding Fock space 
can be obtained from that of eq.(\ref{1}) via a simple projection 
operation.

Let us start by recalling the Fock space of the system (\ref{1}) 
\cite{12,13}. 
According to standard quantum field theories, the space of single particle 
states is a Hilbert space which is now identified with the space
${\cal H} \equiv L^2({\bf R}^s,~{\rm d}^s x)$ of square integrable 
functions on the Euclidean space ${\bf R}^s$ with respect to the 
measure ${\rm d}^s x$. The $n$-particle space ${\cal H}^n$ is 
identified with ${\cal H}^{\otimes n}$, and the Fock space 
${\cal F}({\cal H})$ of the system (\ref{1}) is just the direct 
sum of all the multi-particle subspaces, namely

\be
{\cal F}({\cal H}) = \bigotimes_{n=0}^\infty ~{\cal H}^n,~~~
{\cal H}^0 \equiv {\bf C}. \label{5}
\ee

\noindent Points of the Fock space ${\cal F}({\cal H})$ are denoted by 
their ``coordinates'', $\varphi = (\varphi^{(0)},~\varphi^{(1)},...
~\varphi^{(n)},...)$, where $~\varphi^{(n)} \in {\cal H}^n$. The finite
particle subspace ${\cal F}^0 ({\cal H})$ of ${\cal F}({\cal H})$ is 
consisted of those points for which $\varphi^{(n)}$ is zero for $n$ large
enough.

Consider the subset ${\cal D}^n$ of decomposible vectors in ${\cal H}^n$,

\bd
{\cal D}^n = \{f_1 \otimes f_2 \otimes ... \otimes f_n |f_i \in {\cal H} \}
\subset {\cal H}^n.
\ed

\noindent Define the operations 

\bd
b(f): {\cal D}^n \rightarrow {\cal D}^{n-1},~(n \geq 1);~~~
b^\ast (f): {\cal D}^n \rightarrow {\cal D}^{n+1},~(n \geq 0),~~~
f \in {\cal H}
\ed

\noindent on ${\cal D}^n$ such that  

\bea
& &b(f)~f_1 \otimes ... \otimes f_n 
= \sqrt{n}~(f,~f_1)~f_2 \otimes ... \otimes f_n,\nonumber\\
& &b^\ast(f)~f_1 \otimes ... \otimes f_n 
= \sqrt{n+1}~f \otimes ~f_1 \otimes ... \otimes f_n, \label{6}
\eea

\noindent where $(f,~g) \equiv \int~{\rm d}^s x~\bar{f}(x)~g(x)$, 
$\bar{f}(x)$ represent the conjugation of $f(x)$.

The actions of $b(f)$ and $b^\ast(x)$ can be extended onto the
dense subspace ${\cal L}({\cal D}^n)$ of ${\cal H}$ 
which consists of all the finite 
linear combinations of the elements of ${\cal D}^n$ by linearity. Thus for
any $\varphi^{(n)} \in {\cal L}({\cal D}^n)$, we have

\bea
& & [ b(f)~\varphi ]^{(n)}(x_1,~x_2,~...,~x_n) = \sqrt{n+1} \int 
 {\rm d}^s x~\bar{f}(x)~\varphi^{(n+1)} (x,~x_1,~...,~x_n),
 \nonumber\\
& & [ b^\ast (f)~\varphi ]^{(n)}(x_1,~x_2,~...,~x_n) = \frac{1}{\sqrt{n}} 
 \sum_{k=1}^n~f(x_k)~\varphi^{(n-1)} (x_1,...~\hat{x}_k,~...,~x_n),
 \label{7}
\eea

\noindent where $\hat{x}_k$ indicates where $x_k$ should not appear. One can 
introduce distributions corresponding to the operators $b(f)$ and 
$b^\ast(f)$ respectively as

\be
b(f) = \int {\rm d}^s x~\bar{f}(x)~b(x),~~~~
b^\ast(f) = \int {\rm d}^s x~f(x)~b^\ast(x). \label{8}
\ee

\noindent It can then be shown using eq.(\ref{7}) 
that the distributions $b(x)$ and $b^\ast(x)$ satisfy eq.(\ref{1}).

In order to construct an oscillator algebra which satisfy the reflection 
boundary conditions (\ref{2}-\ref{4}) let us define the operators 
$\pi_n~(n \geq 1)$ acting on ${\cal L}({\cal D}^n)$ as

\be
 [ \pi_n~\varphi ]^{(n)}(x_1,~x_2,~...,~x_n) 
 = K(x_n)~\varphi^{(n)} (x_1,~...,~x_{n-1},~\sigma(x_n)). \label{9}
\ee

\noindent Thanks to (\ref{4}), we have 

\be
\pi_n^2 = Id, \label{10}
\ee

\noindent and in order that $\pi_n$ be Hermitian we also require 

\be
\bar{K}(x) = K(\sigma(x)). \label{11}
\ee

It is obvious that 

\be
P_B^{(n)} = \frac{1}{2} (Id + \pi_n),~~~\bar{P}_B^{(n)} 
= \frac{1}{2} (Id - \pi_n) \label{12}
\ee

\noindent are both projection operators which are orthorgnal to each other.
When supplied with $P_B^{(0)} = Id$, one can extend the set 
$\{ P_B^{(n)} ~|~(n \geq 0) \}$ onto the whole Fock space 
${\cal F}({\cal H})$, 

\be
{\cal F}_B({\cal H}) \equiv P_B~{\cal F}({\cal H}),~~~
P_B |_{{\cal H}^n} = P_B^{(n)}. \label{a1}
\ee

The projected Fock space ${\cal F}_B({\cal H})$ has a natural decomposition
${\cal F}_B({\cal H})=\bigoplus_{n=0}^\infty~{\cal H}_B^{n}$, where 
${\cal H}_B^{n} = P_B^{(n)}~{\cal H}^n$. Correspondingly, 
the projected creation and anihilation operators read

\be
a^\#(f) = P_B~b^\#(f)~P_B, \label{13}
\ee

\noindent where $b^\#$ stands for both $b$ and $b^\ast$. 

By straightforward calculations using eqs.(\ref{7}), (\ref{9}), 
(\ref{12}) and (\ref{13}) we have, 
for arbitrary $\varphi^{(n)} \in {\cal L}({\cal D}_B^{n})$,

\bea
& & [ a(f)~\varphi ]^{(n)}(x_1,~...,~x_n) = \sqrt{n+1}~\int~{\rm d}^s x~
\bar{f}(x)~\varphi^{(n+1)}(x,~x_1,~...,~x_n), \label{14}\\
& & [ a^\ast(f)~\varphi ]^{(n)}(x_1,~...,~x_n) = \frac{1}{\sqrt{n}}
~\sum_{k=1}^n~\left[ f(x_k) +K(x_k)f(\sigma(x_k)) \right] \nonumber\\
& &~~~~\times ~\varphi^{(n-1)}(x_1,~...,\hat{x}_k,~...,~x_n). \label{15}
\eea

\noindent Now define

\bd
a(f) = \int {\rm d}^s x~\bar{f}(x)~a(x),~~~~
a^\ast(f) = \int {\rm d}^s x~f(x)~a^\ast(x) 
\ed

\noindent in analogy to (\ref{8}), we have, from (\ref{14}-\ref{15}),

\bea
& & [ a(x),~a(y) ] = [ a^\ast(x),~ a^\ast(y) ] = 0,\nonumber\\
& & [ a(x),~a^\ast(y) ] = \delta(y-x) + K(x) \delta(y -\sigma(x)).
\label{16}
\eea

\noindent This algebra is just what we have called the oscillator algebra
satisfying the reflection boundary condition. Since the vacuum state
$\Omega_B$ of this algebra is just the image of that of the algebra 
(\ref{1}) under the projection (\ref{a1}), 
and the vacuum state $\Omega$ of (\ref{1}) belongs to 
${\cal H}^0$ and $P_B^{(0)} = Id$, we have the conclusion $\Omega_B 
= \Omega$. 

Let us end this section by proving that the algebra (\ref{16}) 
indeed satisfy the 
reflection boundary conditions (\ref{2}-\ref{4}). Since 
$a^\ast(f) \Omega_B \in {\cal H}_B = P_B^{(1)} {\cal H}$, we have

\be
\bar{P}^{(1)}_B(a^\ast(f) \Omega_B) = 0, \label{17}
\ee

\noindent or written as

\be
a^\ast(f) \Omega_B - \pi_1a^\ast(f) \Omega_B = 0. \label{18}
\ee

\noindent Using the definition of $\pi_1$ and rewriting $a^\ast(f)$ 
in terms of the distribution $a^\ast(x)$, one can easily see that 
eq.(\ref{18}) is equivalent to (\ref{2}), and we thus have obtained the 
oscillator algebra (\ref{17}) satisfying the reflection boundary conditions 
(\ref{2}-\ref{4}) and its Fock space ${\cal F}_B({\cal H})$.

\section{Oscillator algebra with reflection boundary condition 
and generalized statistics}
\setcounter{equation}{0}

In the last section we obtained the oscillator algebra (\ref{16}) which 
satisfy the reflection boundary conditions (\ref{2}-\ref{4}). However, 
this algebra still obeys the usual bose statistics, i.e. $a(x)$ and $a(y)$
exchange in the way the usual bosons do. In this section we shall 
generalize the construction to include the generalized statistics. For this
purpose we adopt the technique introduced in Ref.\cite{12,13}. On arbitrary
$\varphi^{(n)} \in {\cal H}^n$ we introduce the following operations,

\bea
& & [ s_i \varphi ]^{(n)}(x_1,~...,~x_n) = R(x_i,~x_{i+1})
~\varphi^{(n)}(x_1,~...,~x_{i+1},~x_i,~...,~x_n),\nonumber\\
& & R(x_i,~x_{i+1}) R(x_{i+1},~x_{i}) = 1,~~~~
\bar{R}(x_i,~x_{i+1})= R(x_{i+1},~x_{i}), \label{19}\\
& &(i=1,~...,~n-1). \nonumber
\eea

\noindent It is an easy practice to prove that $s_i~(i=1,~...,~n-1)$ and 
$s_n \equiv \pi_n$ satisfy the following relations

\bea
& &(s_i s_j)^{m_{ij}} = 1,\nonumber\\
& &m_{ij}=\left\{
\begin{array}{ll}
1, & $$ |i-j| = 0$$\cr
2, & $$ |i-j| > 1$$\cr
3, & $$ |i-j| = 1~~{\rm and}~~i,~j \neq n $$\cr
4, & $$ |i-j| = 1~~{\rm and}~~max(i,~j)=n $$
\end{array}
\label{20}
\right.
\eea

\noindent provided the ``exchange factor'' $R(x_i,~x_{i+1})$ satisfy the 
following condition

\be
R(x,~y) R(y,~\sigma(x)) = R(x,~\sigma(y)) R(\sigma(y),~\sigma(x)).
\label{21}
\ee

\noindent Eq.(\ref{20}) is exactly the generating relation of the Weyl group
${\cal B}_n$ which corresponds to the Lie algebra $B_n$ and $C_n$ \cite{17}.
Therefore, eqs.(\ref{9}) and (\ref{19}) actually define an action of the 
Weyl group ${\cal B}_n$ on ${\cal H}^n$. Eq.(\ref{21}) may seem to be 
misterious at a first glance. However, it will be clear in the next section
that this is actually the most degenerated case of the (generalized) 
boundary Yang-Baxter equation.

Our purpose is to find a subspace ${\cal F}_R({\cal H})$ of 
${\cal F}({\cal H})$ on which the exchange of any two neighboring 
particles sitting for example at cites $i$ and $i+1$ in an arbitrary 
$n$-particle state give rise to the ``exchange factor'' 
$R(x_i,~x_{i+1})$, and the single particle state still satisfy 
the conditions (\ref{2}-\ref{4}). To achieve this we now define 
the projection operators

\be
P_R^{(n)} = \frac{1}{{\rm dim}{\cal B}_n} \sum_{g \in {\cal B}_n} ~g,~~~
{n \geq 2} \label{22}
\ee

\noindent and also let $P_R^{(0)} =Id,~P_R^{(1)}=P_B^{(1)}$. It follows 
from the fact $\pi_n \in {\cal B}_n$ and the rearrangement theorem
that $(Id - \pi_n) P_R^{(n)} = 0$, which implies that $P_R^{(n)}$ and 
$\bar{P}_B^{(n)}$ are orthogonal to each other. Similar to the case 
of $P_B^{(n)}$, one can also extend the action of $P_R^{(n)}$ to the 
whole Fock space ${\cal F}({\cal H})$,

\bd
{\cal F}_R({\cal H}) = P_R {\cal F}({\cal H}),
~~~P_R|_{{\cal H}^n} = P_R^{(n)}.
\ed

\noindent Correspondingly, the operators $b^\#$ are projected to $a_R^\#$,

\bd
a_R^\# = P_R b^\# P_R.
\ed

It can be proved that the action of $a_R^\#$ on arbitrary $\varphi^{(n)}
\in {\cal L}({\cal D}_R^n)$ gives rise to the following results,

\bea
& &[ a_R(f) \varphi ]^{(n)}(x_1,~...,~x_n) = \sqrt{n+1}~\int~{\rm d}^s x~
\bar{f}(x)~\varphi^{(n+1)}(x,~x_1,~...,~x_n),\nonumber\\
& &[ a^\ast_R(f) \varphi ]^{(n)}(x_1,~...,~x_n) = \frac{1}{\sqrt{n}}
\left[ R(x_{n-1},~x_n)...R(x_1,~x_n)f(x_n) \right.\nonumber\\
& & ~~~~\left.+ K(x_n) R(x_{n-1},~\sigma(x_n))...
R(x_1,~\sigma(x_n))f(\sigma(x_n)) \right]
\varphi^{(n-1)}(x_{1},~...,~x_{n-1}) \nonumber\\
& &~~~~+ \frac{1}{\sqrt{n}} \sum_{k=1}^{n-1}~R(x_k,~x_{k+1})...R(x_k,~x_n)
\left[ R(x_{n},~x_k)...R(x_{k+1},~x_k)\right.\nonumber\\
& &~~~~\times~R(x_{k-1},~x_k)...R(x_{1},~x_k) f(x_k) 
 + K(x_k) R(x_{n},~\sigma(x_k))...R(x_{k+1},~\sigma(x_k)) 
\nonumber\\
& &~~~~\left. \times R(x_{k-1},~\sigma(x_k))...
R(x_{1},~\sigma(x_k)) f(\sigma(x_k)) \right]
\varphi^{(n-1)}(x_{1},~...,\hat{x}_k,~...,~x_{n}).
\label{23}
\eea

\noindent After rather tedious calculations, one can get

\bea
& & a_R(x) a_R(y) - R(y,~x) a_R(y) a_R(x) = 0, \nonumber\\
& & a^\ast_R(x) a^\ast_R(y) - R(y,~x) a^\ast_R(y) a^\ast_R(x) = 0, 
\label{24}\\
& & a_R(x) a^\ast_R(y) - R(x,~y) a^\ast_R(y) a_R(x) = \delta(y-x) +
F(x) \delta(y-\sigma(x)),\nonumber
\eea

\noindent where $F(x)$ is an Hermitian operator acting on
${\cal F}_R({\cal H})$ as

\be
 [ F(x) \varphi]^{(n)} (x_1,~...,~x_n) = 
R(x,~x_1)...R(x,~x_n) K(x) R(x_{n},~\sigma(x))...R(x_1,~\sigma(x))
\varphi^{(n)}(x_1,~...,~x_n). \label{25}
\ee

\noindent Using eqs.(\ref{23}) and (\ref{25}) one can also obtain 
the exchange relations

\bea
& &R(x,~y) F(x) a_R(y) R(y,~\sigma(x)) - a_R(y) F(x) = 0, \nonumber\\
& &F(x) a_R^\ast(y) - R(y,~x) a_R^\ast(y) F(x) R(\sigma(x),~y)= 0, 
\label{26}\\
& &F(x) F(y) - F(y) F(x) = 0. \nonumber
\eea

\noindent Eqs.(\ref{24}) and (\ref{26}) together form an oscillator algebra
satisfying the reflection boundary conditions (\ref{2}-\ref{4}) and obeying 
the $R$-generalized statistics. Notice that if in eq.(\ref{20}) we did not
include $\pi_n$ the corresponding generating relations would become that of 
the permutation group, and the resulting algebra will turn out to be the 
oscillator algebra without the reflection condition \cite{12,13}.

\section{Multicomponent generalization (summary)}
\setcounter{equation}{0}

In this section we shall generalize the construction even further to the case
when more than one kinds of creation and anihilation operators are involved.
Since the multicomponent case is far more complicated comparing to the 
single component case and such case is believed to be very important in 
the context of quantum factorizable scattering theory, we shall give the 
detailed calculations elsewhere and present here only with the brief summary.

Suppose there are $N$-kinds of elementary excitations in the 
system. Then the space of single-particle states before the projection  
will be

\be
{\cal H} = \bigoplus_{\alpha=1}^{N}~L^2({\bf R}^s,~{\rm d}^s x). \label{27}
\ee

\noindent Correspondingly the Fock space reads

\be
{\cal F}({\cal H}) = \bigoplus_{n=0}^\infty {\cal H}^n. \label{28}
\ee
\noindent In ${\cal H}$ the inner product is defined as

\be
(f,~g) = \int {\rm d}^s x ~f^{\dagger \alpha} (x) g_{\alpha}(x) 
= \sum_{\alpha=1}^{N} \int {\rm d}^s x~\bar{f}_\alpha (x) g_\alpha(x),
\label{29}
\ee

\noindent where $f=(f_1,~...,~f_N)^T \in {\cal H}$. In the above system 
we introduce the $N^2 \times N^2$ matrix 
$\left(R_{\alpha_1 \alpha_2}^{\beta_1 \beta_2} (x_1,~x_2) \right)$ as
exchange factor and the $N \times N$ matrix 
$\left(K_\alpha^\beta(x)\right)$ as reflection factor, which satisfy 
the following constraints,

\bea
& &R_{\alpha_1 \alpha_2}^{\epsilon_1 \epsilon_2}(x_1,~x_2)
R_{\epsilon_2 \alpha_3}^{\gamma_2 \beta_3}(x_1,~x_3)
R_{\epsilon_1 \gamma_2}^{\beta_1 \beta_2}(x_2,~x_3)\nonumber\\
& &~~~~~= R_{\alpha_2 \alpha_3}^{\epsilon_2 \epsilon_3}(x_2,~x_3)
R_{\alpha_1 \epsilon_2}^{\beta_1 \gamma_2}(x_1,~x_3)
R_{\gamma_2 \epsilon_3}^{\beta_2 \beta_3}(x_1,~x_2),\label{30}\\
& &R_{\alpha_1 \alpha_2}^{\gamma_1 \gamma_2}(x_1,~x_2)
R_{\gamma_1 \gamma_2}^{\beta_1 \beta_2}(x_2,~x_1)
= \delta_{\alpha_1}^{\beta_1} \delta_{\alpha_2}^{\beta_2},\label{31}\\
& &\bar{R}_{\alpha_1 \alpha_2}^{\beta_1 \beta_2}(x_1,~x_2)
=R_{\beta_1 \beta_2}^{\alpha_1 \alpha_2}(x_2,~x_1),\label{32}\\
& &R_{\alpha_1 \alpha_2}^{\gamma_1 \gamma_2}(x_1,~x_2)
K_{\gamma_2}^{\delta_2}(x_1) 
R_{\gamma_1 \delta_2}^{\beta_1 \rho_2}(x_2,~\sigma(x_1))
K_{\rho_2}^{\beta_2}(x_2)\nonumber\\
& &~~~~= K_{\alpha_2}^{\gamma_2}(x_2) 
R_{\alpha_1 \gamma_2}^{\gamma_1 \delta_2}(x_1,~\sigma(x_2))
K_{\delta_2}^{\rho_2}(x_1) 
R_{\gamma_1 \rho_2}^{\beta_1 \beta_2}(\sigma(x_2),~\sigma(x_1)),\label{33}\\
& &K_{\gamma}^{\alpha}(x) K_{\beta}^{\gamma}(\sigma(x)) 
= \delta_{\beta}^{\alpha},\label{34}\\
& &\bar{K}_{\beta}^{\alpha}(x)= K_{\alpha}^{\beta}(\sigma(x)).\label{35}
\eea

\noindent These constraints are generalizations of the well-known
(braid type) Yang-Baxter equation, the unitarity and cross unitarity 
conditions for the $R$-matrix, the boundary Yang-Baxter equation and the 
unitarity and cross unitarity conditions of the $K$-matrix. 
Since we did not introduce an explicit conjugation matrix, the presentation
of eqs.(\ref{32}) and (\ref{35}) are slightly different from the form of
the usual cross unitarity conditions for the $R$- and $K$-matrices. 
Notice that the counterpart of eq.(\ref{32}) did not appear in the last 
section because it holds trivially there. Notice also that eq.(\ref{35}) 
reduces to eq.(\ref{21}) in the one component case.

Using the above equations one can show that the operators $s_i,~(i=1,~
...,~n-1)$ and $s_n \equiv \pi_n$, defined as follows, satisfy eq. (\ref{20}),

\bee
& & [ s_i \varphi ]^{(n)}_{\alpha_1 ... \alpha_n} (x_1,~...,~x_n) \\
& &~~~~= 
[ R_{i,i+1}(x_i,~x_{i+1}) ]_{\alpha_1 ... \alpha_n}^{\beta_1 ... \beta_n}
\varphi_{\beta_1 ... \beta_n}^{(n)}(x_1,~...,~x_{i+1},~x_i,~...,~x_n),
~~(n \geq 2) \\
& & [ \pi_n \varphi ]^{(n)}_{\alpha_1 ... \alpha_n} (x_1,~...,~x_n) 
= [ K_n(x_n)]_{\alpha_1 ... \alpha_n}^{\beta_1 ... \beta_n}
\varphi_{\beta_1 ... \beta_n}^{(n)}(x_1,~...,~x_{n-1},~\sigma(x_n)),
~~(n \geq 1),\\
& & [ R_{i,j}(x_i,~x_j) ]_{\alpha_1 ... \alpha_n}^{\beta_1 ... \beta_n}
\equiv \delta_{\alpha_1}^{\beta_1}...\widehat{\delta_{\alpha_i}^{\beta_i}}
...\widehat{\delta_{\alpha_j}^{\beta_j}}...\delta_{\alpha_n}^{\beta_n}
R_{\alpha_i \alpha_j}^{\beta_i \beta_j}(x_i,~x_j),\\
& &[ K_{j}(x_j) ]_{\alpha_1 ... \alpha_n}^{\beta_1 ... \beta_n}
\equiv \delta_{\alpha_1}^{\beta_1}...
\widehat{\delta_{\alpha_j}^{\beta_j}}...\delta_{\alpha_n}^{\beta_n}
K_{\alpha_j}^{\beta_j}(x_j).
\eee

\noindent Therefore, following exactly the parallel procedure as in the 
last section, one can define the projection operations and obtain the 
projected Fock space. Finally, in the projected Fock space, one can obtain 
a complete set of creation operators and anihilation operators which 
obey the following algebra,

\bea
& &a_\alpha(x)a_\beta(y) - R_{\beta \alpha}^{\delta \gamma}(y,~x) 
a_\gamma(y) a_\delta(x) = 0,\label{36}\\
& &a^{\ast \alpha}(x)a^{\ast \beta} (y) - a^{\ast \gamma}(y) 
a^{\ast \delta} (x) R^{\alpha \beta}_{\gamma \delta}(y,~x)  = 0,\label{37}\\
& &a_\alpha(x) a^{\ast \beta}(y) - a^{\ast \gamma}(y)
R_{\alpha \gamma}^{\beta \delta}(x,~y) a_\delta(x)
 = \delta_{\alpha}^{\beta} \delta(y-x) + F_{\alpha}^{\beta} (x) 
\delta(y-\sigma(x)),\label{38}\\
& &R_{\alpha \gamma}^{\epsilon \tilde{\alpha}}(x,~y)
F_{\tilde{\alpha}}^{\tilde{\beta}}(x) a_{\tilde{\gamma}}(y) 
R_{\epsilon \tilde{\beta}}^{\beta \tilde{\gamma}}(y,~\sigma(x))
- a_{\gamma}(y) F_{\alpha}^{\beta}(x) = 0,\label{39}\\
& &F_{\beta}^{\alpha}(x) a^{\ast \gamma} (y) - 
R_{\beta \tilde{\gamma}}^{\epsilon \tilde{\beta}}(y,~x)
a^{\ast \tilde{\gamma}} (y) F_{\tilde{\beta}}^{\tilde{\alpha}}(x) 
R_{\epsilon \tilde{\alpha}}^{\alpha \gamma}(\sigma(x),~y)
 = 0,\label{40}\\
& &R_{\alpha_1 \alpha_2}^{\gamma_1 \gamma_2}(x,~y)
F_{\gamma_2}^{\delta_2}(x) 
R_{\gamma_1 \delta_2}^{\beta_1 \rho_2}(y,~\sigma(x))
F_{\rho_2}^{\beta_2}(y)\nonumber\\
& &~~~~= F_{\alpha_2}^{\gamma_2}(y) 
R_{\alpha_1 \gamma_2}^{\gamma_1 \delta_2}(x,~\sigma(y))
F_{\delta_2}^{\rho_2}(x) 
R_{\gamma_1 \rho_2}^{\beta_1 \beta_2}(\sigma(y),~\sigma(x)),\label{41}
\eea

\noindent where $F_\alpha^\beta(x)$ is defined as

\bee
& & [ F_\alpha^\beta(x) \varphi ]^{(n)}_{\alpha_1 ...\alpha_n}
(x_1,~...,~x_n) \\
& &~~~~= [ R_{01}(x,~x_1)...R_{n-1,~n}(x,~x_n)K_n(x)
R_{n-1,~n}(x_n,~\sigma(x))...
R_{01}(x_1,~\sigma(x))]_{\alpha \alpha_1 ...\alpha_n}
^{\beta \beta_1 ... \beta_n}\\
& &~~~~\times \varphi^{(n)}_{\beta_1 ... \beta_n}(x_1 ... x_n).
\eee

\noindent Eqs.(\ref{36}-\ref{40}) form an multicomponent generalization 
of the algebra obtained in the last section. Notice that if 
$F_\alpha^\beta(x) = 0$, then the above algebta will become the 
well-known Faddeev-Zamolodchikov algebra appeared in the context of 
quantum factorizable scattering theory. In the present case
$a^{\ast\alpha}(x)$ obeys the reflection boundary condition

\bd
a^{\ast \alpha}(x) \Omega_B = K_\beta^\alpha(x) 
a^{\ast \beta}(\sigma(x)) \Omega_B.
\ed

\noindent This observation may imply the possible application of the algebra
(\ref{36}-\ref{41}) in the boundary factorizable scattering theories.

\section{Discussions}

In the above we have obtained the oscillator algebra and the corresponding 
Fock space which satisfy the reflection boundary condition and obey the 
generalized statistics. These results already enables us to calculate the 
$n$-particle correlation functions in the corresponding system. For example,
in the one component case, the $n$-particle states are nothing but the 
linear combinations of the vectors of the form 
$a^\ast_R(x_1)...a^\ast_R(x_n)\Omega_B$. Therefore the $n$-point correlation
function can be expressed as

\bd
\omega_n(x_1,~...,~x_n;~y_1,~...,~y_n)=
\frac{(a^\ast_R(x_1)...a^\ast_R(x_n)\Omega_B,~a^\ast_R(y_1)...
a^\ast_R(y_n)\Omega_B)}{(\Omega_B,~\Omega_B)}.
\ed

\noindent Actually one can write out an iterative relation for $\omega_n$,

\bee
& &\omega_n(x_1,~...,~x_n;~y_1,~...,~y_n)\\
& &~~~~= [ \delta(y_1 - x_1) + R(x_1,~y_2)...R(x_1,~y_n)K(x_1)\\
& &~~~~\times R(y_n,\sigma(x_1))...R(y_2,~\sigma(x_1))
\delta(y_1-\sigma(x_1)) ]\\
& &~~~~\times \omega_{n-1}(x_2,~...,~x_n;~y_2,~...,~y_n)\\
& &~~~~+ \sum_{k=2}^{n} R(x_1,~y_1)...R(x_1,~y_{k-1})
[ \delta(y_k - x_1) + R(x_1,~y_{k+1})...R(x_1,~y_n)K(x_1)\\
& &~~~~\times 
R(y_n,\sigma(x_1))...R(y_{k+1},~\sigma(x_1))\delta(y_k-\sigma(x_1)) ]\\
& &~~~~ \times \omega_{n-1}(x_2,~...,~x_n;~y_1,~...,\hat{y}_k,~...,~y_n).
\eee

\noindent The first few of the $\omega_n$ read

\bee
& &\omega_0 = 1,~~~~\omega_1(x,~y) =\delta(y-x) + K(x)\delta(y-\sigma(x)),\\
& &\omega_2(x_1,~x_2;~y_1,~y_2) = [ \delta(y_1- x_1) 
+ R(x_1,~y_2)K(x_1)R(y_2,~\sigma(x_1)) \delta(y_1-\sigma(x_1))] \\
& &~~~~\times [\delta(y_2-x_2) + K(x_2)\delta(y_2-\sigma(x_2)) ]\\
& &~~~~+R(x_1,~y_1) [ \delta(y_2- x_1) 
+ K(x_1)\delta(y_2-\sigma(x_1))] \\
& &~~~~\times [ \delta(y_2 -x_2) + K(x_2)\delta(y_2-\sigma(x_2)) ].
\eee

Many problems related to the construction in this article are still
left for study. For examples, the second quantization based on the 
oscillator algebra of this article, the coherent states, the partition
function of the multi-particle system and the properties of the ideal gas
consisted of the (quasi-)particles all deserves to be further studied.
Particularly, for the second quantization problem, Liguori and Mintchev
\cite{12,13} have considered the possibility of obtaining multi-particle
Hamiltonian with nontrivial interactions starting from a single free 
particle Hamiltonian in the case when there is no boundary reflection and 
successfully applied the results to the Leinaas-Myrheim anion system 
\cite{18,19}. It is desirable that the similar construction can be carried 
out in the boundary case and the resulting Hamiltonians will include not only
interactions but also self-interactions. We would like to come 
back to this point at a later time.

\section*{Acknowledgement}
The author is grateful to Bo-yu Hou, Pei Wang, Kang-jie Shi, Wen-li Yang
for discussions. He would also like to thank A. Liguori for explaining their 
work on the boundaryless case. This work is supported in part by the National  
Natural Science Found od China.


\newpage


\begin{thebibliography}{33}
\bibitem{1} A. J. Macfarlane, J. Phys. A: Math. Gen. 22 (1989) 4581.

\bibitem{2} L. C. Biedenharn, J. Phys. A: Math. Gen. 22 (1989) L873.

\bibitem{3} O. W. Greenberg, Phys. Rev. Lett. 64 (1990) 705; Phys. Rev. D43
(1991) 4111.

\bibitem{4} R. N. Mohapatra, Phys. Lett. B242 (1990) 407.

\bibitem{5} R. Chakrabarti, R. Jagannathan, J. Phys. A: Math. Gen. 
24 (1991) L711.

\bibitem{6} G. Brodimas, A. Jannussis, R. Mignani, J. Phys. A: Math. Gen. 
25 (1992) L329.

\bibitem{7} M. Arik, E. Demircan, T. Turgut, L. Ekinci, M. Mungan, Z. Phys. 
C: Particle \& Fields 55 (1992) 89.

\bibitem{8} V. Kalmeyer, R. B. Laughlin, Phys. Rev. Lett. 59 (1987) 2095.

\bibitem{9} S. Forte, Rev. Mod. Phys. 64 (1992) 193.

\bibitem{10} B. I. Halperin, Phys. Rev. Lett.52 (1984) 1583.

\bibitem{11} R. B. Laughlin, Phys. Rev. Lett. 50 (1983) 1395.

\bibitem{12} A. Liguori, M. Mintchev, Commun, Math. Phys. 169 (1995) 635.

\bibitem{13} A. Liguori, M. Mintchev, Lett. Math. Phys. 33 (1995) 283.

\bibitem{14} S. Ghoshal, A. Zamolodchkov, Int. J. Mod. Phys. A9 (1994) 3841.

\bibitem{15} M. Jimbo, R. Kedom, T. Kojima, H. Konno, T. Miwa, Nucl. Phys. 
B441 (1995) 437.

\bibitem{16} B.-Y. Hou, K.-J. Shi, Y.-S. Wang, W.-L. Yang, NWU-IMP preprint 
1995.

\bibitem{17} Finite reflection group, Graduate Text in Mathematics 99,
Spinger-Verlag.

\bibitem{18} J. M. Leinaas, J. Myrheim, Nuovo. Cim. B37 (1977) 1.

\bibitem{19} G. V. Duune, A. Lerda, S. Sciuto, C. A. Trugenberger, 
Nucl. Phys. B370 (1992) 601.

\end{thebibliography}
\end{document}